\documentclass[aps,prl,reprint,twocolumn,groupedaddress,floatfix]{revtex4-1}

\usepackage{amsmath}
\usepackage{graphicx}

\newcommand{\physrep}{Phys.~Rep.} 
\newcommand{\apjl}{Astrophys.~J.~Lett.} 
\newcommand{\apjs}{Astrophys.~J.~Suppl.~Ser.} 
\newcommand{\aap}{Astron.~\&~Astrophys.} 
\newcommand{\araa}{Ann.~Rev.~Astron.~Astrophys.} 
\newcommand{\mnras}{Mon.~Not.~R.~Astron.~Soc.} 

\newcommand{\half}{(1/2)}
\newcommand{\bfu}{\mathbf{u}}
\newcommand{\bfB}{\mathbf{B}}
\newcommand{\bfomega}{\boldsymbol{\omega}}
\newcommand{\vect}[1]{{\mathbf{#1}}}
\newcommand{\cs}{c_\mathrm{s}}
\newcommand{\cm}{\mbox{cm}}
\newcommand{\g}{\mbox{g}}
\newcommand{\G}{\mbox{Gauss}}
\newcommand{\Ma}{\mathcal{M}}
\newcommand{\ted}{t_\mathrm{ed}}
\newcommand{\Em}{E_{\mathrm{m}}}
\newcommand{\Emzero}{E_{\mathrm{m0}}}
\newcommand{\Ek}{E_{\mathrm{k}}}
\newcommand{\satlev}{\left(\Em/\Ek\right)_\mathrm{sat}}
\newcommand{\solratio}{E_\mathrm{sol}/E_\mathrm{tot}}
\newcommand{\Rm}{\mathrm{Rm}}
\newcommand{\Pm}{\mathrm{Pm}}

\begin{document}

\preprint{}
\title{Mach Number Dependence of Turbulent Magnetic Field Amplification: \\Solenoidal versus Compressive Flows}

\author{C.~Federrath}
\email{{\footnotesize federrath@uni-heidelberg.de}}
\altaffiliation{{\footnotesize Ecole Normale Sup\'{e}rieure de Lyon, F-69364 Lyon, France}}
\altaffiliation{{\footnotesize Zentrum f\"ur Astronomie der Universit\"at Heidelberg, D-69120 Heidelberg, Germany}}

\author{G.~Chabrier}
\altaffiliation{{\footnotesize Ecole Normale Sup\'{e}rieure de Lyon, F-69364 Lyon, France}}
\altaffiliation{{\footnotesize School of Physics, University of Exeter, Exeter, EX4 4QL, UK}}

\author{J.~Schober}
\altaffiliation{{\footnotesize Zentrum f\"ur Astronomie der Universit\"at Heidelberg, D-69120 Heidelberg, Germany}}

\author{R.~Banerjee}
\altaffiliation{{\footnotesize Hamburger Sternwarte, D-21029 Hamburg, Germany}}

\author{R.~S.~Klessen}
\altaffiliation{{\footnotesize Zentrum f\"ur Astronomie der Universit\"at Heidelberg, D-69120 Heidelberg, Germany}}

\author{D.~R.~G.~Schleicher}
\altaffiliation{{\footnotesize Georg-August-Universit\"at, Institut f\"ur Astrophysik, D-37077 G\"ottingen, Germany}}

\date{\today}

\begin{abstract}
We study the growth rate and saturation level of the turbulent dynamo in magnetohydrodynamical simulations of turbulence, driven with solenoidal (divergence-free) or compressive (curl-free) forcing. For models with Mach numbers ranging from 0.02 to 20, we find significantly different magnetic field geometries, amplification rates, and saturation levels, decreasing strongly at the transition from subsonic to supersonic flows, due to the development of shocks. Both extreme types of turbulent forcing drive the dynamo, but solenoidal forcing is more efficient, because it produces more vorticity.
\end{abstract}


\maketitle

The turbulent dynamo is the most important process to amplify a small initial magnetic field \cite[][]{BrandenburgSubramanian2005}. The growth of the magnetic field is exponential, which leads to dynamically significant magnetic energies on short time scales. Dynamo action ranges from the Earth and the Sun \cite{CattaneoHughes2001}, over the interstellar medium to whole galaxies \cite{BeckEtAl1996}. Although the physical conditions (e.g., the different compressibility of the plasmas) and flow geometries are extremely different across these objects, dynamo action has been confirmed in all of them. For instance, in the Earth and the Sun, the dynamo is driven by subsonic flows. In contrast, interstellar clouds and galaxies are dominated by highly supersonic, compressible turbulence.

The main objective of this Letter is to investigate fundamental properties of turbulent dynamo amplification of magnetic fields by making systematic numerical experiments, in which we can control the compressibility of the plasma by varying the Mach number and the energy injection mechanism (forcing) of the turbulence. We consider flows with Mach numbers ranging from $\Ma=0.02$ to $20$, covering a much larger range than in any previous study. Haugen et al.~\cite{HaugenBrandenburgMee2004} provided critical Reynolds numbers for dynamo action, but did not investigate growth rates or saturation levels, and studied only $0.1\leq\Ma\leq2.6$. The energy released by, e.g., supernova explosions, however, drives interstellar and galactic turbulence with Mach numbers up to 100 \cite{MacLowKlessen2004}. Thus, much higher Mach numbers have to be investigated. It is furthermore tempting to associate such supernova blast waves with compressive forcing of turbulence \cite{MeeBrandenburg2006,SchmidtFederrathKlessen2008,FederrathDuvalKlessenSchmidtMacLow2010}. Mee \& Brandenburg \cite{MeeBrandenburg2006} concluded that it is very hard to excite the turbulent dynamo with such curl-free forcing, because vorticity is not directly injected. In this Letter, we show that the turbulent dynamo is driven by curl-free injection mechanisms, and quantify the amplification as a function of compressibility of the plasma. This is the first study--to the best of our knowledge--addressing the Mach number and forcing dependence of the turbulent dynamo in detail. The main questions addressed are: How does the turbulent dynamo depend on the Mach number of the flow? What are the growth rates and saturation levels in the supersonic and subsonic regimes of turbulence? What is the field geometry and amplification mechanism?

To address these questions, we compute numerical solutions of the compressible, nonideal, three-dimensional, magnetohydrodynamical (MHD) equations with the grid code FLASH \cite{FryxellEtAl2000},
\begin{equation}
\def\arraystretch{1.2}
\begin{array}{@{}l@{}}
\partial_t \rho + \nabla\cdot\left(\rho \bfu\right)=0 ,\\
\partial_t\!\left(\rho \bfu\right) + \nabla\cdot\left(\rho \bfu\!\otimes\!\bfu - \bfB\!\otimes\!\bfB\right) + \nabla p_\star = \nabla\cdot\left(2\nu\rho\boldsymbol{\mathcal{S}}\right) + 
\rho{\bf F}, \\
\partial_t E + \nabla\cdot\left[\left(E+p_\star\right)\bfu - \left(\bfB\cdot\bfu\right)\bfB\right] = \\
\multicolumn{1}{r}{\nabla\cdot\left[2\nu\rho\bfu\cdot\boldsymbol{\mathcal{S}}+\bfB\times\left(\eta\nabla\times\bfB\right)\right],}\\
\partial_t \bfB = \nabla\times\left(\bfu\times\bfB\right) + \eta\nabla^2\bfB, \\
\nabla\cdot\bfB = 0,
\end{array}
\label{eq:mhd}
\end{equation}
where $\rho$, $\bfu$, $p_\star=p+ \half\left|\bfB\right|^2$, $\bfB$, and $E=\rho \epsilon_\mathrm{int} + \half\rho\left|\bfu\right|^2 + \half\left|\bfB\right|^2$ denote density, velocity, pressure (thermal and magnetic), magnetic field, and total energy density (internal, kinetic, and magnetic). Viscous interactions are included via the traceless rate of strain tensor, $\mathcal{S}_{ij}=(1/2)(\partial_i u_j+\partial_j u_i)-(1/3)\delta_{ij}\nabla\cdot\bfu$, and controlled by the kinematic viscosity, $\nu$. We also include physical diffusion of $\bfB$, which is controlled by the magnetic diffusivity, $\eta$. The MHD equations are closed with a polytropic equation of state, $p=\cs^2\rho$, such that the gas remains isothermal with constant sound speed $\cs$. To drive turbulence with a given Mach number, we apply the forcing term ${\bf F}$ as a source term in the momentum equation. The forcing is modeled with a stochastic Ornstein-Uhlenbeck process \cite{SchmidtEtAl2009,FederrathDuvalKlessenSchmidtMacLow2010}, such that ${\bf F}$ varies smoothly in space and time with an autocorrelation equal to the eddy-turnover time, $\ted=L/(2\Ma\cs)$ at the largest scales, $L/2$ in the periodic simulation domain of size $L$. $\Ma=u_\mathrm{rms}/\cs$ denotes the root-mean-squared (rms) Mach number, the ratio of rms velocity and sound speed. The forcing is constructed in Fourier space such that kinetic energy is injected at the smallest wave numbers, $1<\left|\mathbf{k}\right|L/2\pi<3$. We decompose the force field into its solenoidal and compressive parts by applying a projection in Fourier space. In index notation, the projection operator reads
$\mathcal{P}_{ij}^\zeta\,(\vect{k}) = \zeta\,\mathcal{P}_{ij}^\perp+(1-\zeta)\,\mathcal{P}_{ij}^\parallel = \zeta\,\delta_{ij}+(1-2\zeta)\,k_i k_j/|\vect{k}|^2$,
where $\mathcal{P}_{ij}^\perp$ and $\mathcal{P}_{ij}^\parallel$ are the solenoidal and compressive projection operators. This projection allows us to construct a solenoidal (divergence-free) or compressive (curl-free) force field by setting $\zeta=1$ (sol) or $\zeta=0$ (comp).

For most of the simulations, we set the kinematic viscosity $\nu$ and the magnetic diffusivity $\eta$ to zero, and thus solve the ideal MHD equations. In this case, the dissipation of kinetic and magnetic energy is due to the discretization of the fluid equations. However, we did not add any artificial viscosity. Here, we use Riemann solvers, which capture shocks also in the absence of artificial viscosity. In addition to the ideal MHD simulations, however, we also solved the full, nonideal MHD system, Eq.~\ref{eq:mhd}, for four representative models to show that our results are physical and robust against changes in the numerical scheme. For the ideal MHD simulations, we use the positive-definite, split Riemann scheme HLL3R \citep{WaaganFederrathKlingenberg2011} in FLASH v2.5, while our nonideal MHD simulations were preformed with the unsplit staggered mesh scheme in FLASH v4 \citep{LeeDeane2009}, using a third-order reconstruction, constrained transport to maintain $\nabla\cdot\bfB=0$ to machine precision, and the HLLD Riemann solver \citep{MiyoshiKusano2005}. We ran simulations with $128^3$, $256^3$, and $512^3$ grid cells, showing convergence of our results below.

We start our numerical experiments by setting $L=1.24\times10^{19}\,\cm$, uniform $\bfu_0=0$, $\rho_0=1.93\times10^{-21}\,\g\,\cm^{-3}$, $\cs=2\times10^4\,\cm\,\mathrm{s}^{-1}$, and $\bfB=(0,0,B_{0z})$ with $B_{0z}=4.4\times10^{-16}\,\G$ in $z$-direction, corresponding to an extremely high initial plasma $\beta=2p/B^2=10^{20}$. These values are motivated by dynamo studies of primordial clouds \cite{SchleicherEtAl2010,SurEtAl2010,FederrathSurSchleicherBanerjeeKlessen2011}, but in the following, we scale all quantities to dimensionless units to address fundamental questions of magnetic field amplification in compressible plasmas.

\begin{figure}[t]
\centerline{\includegraphics[width=0.99\linewidth]{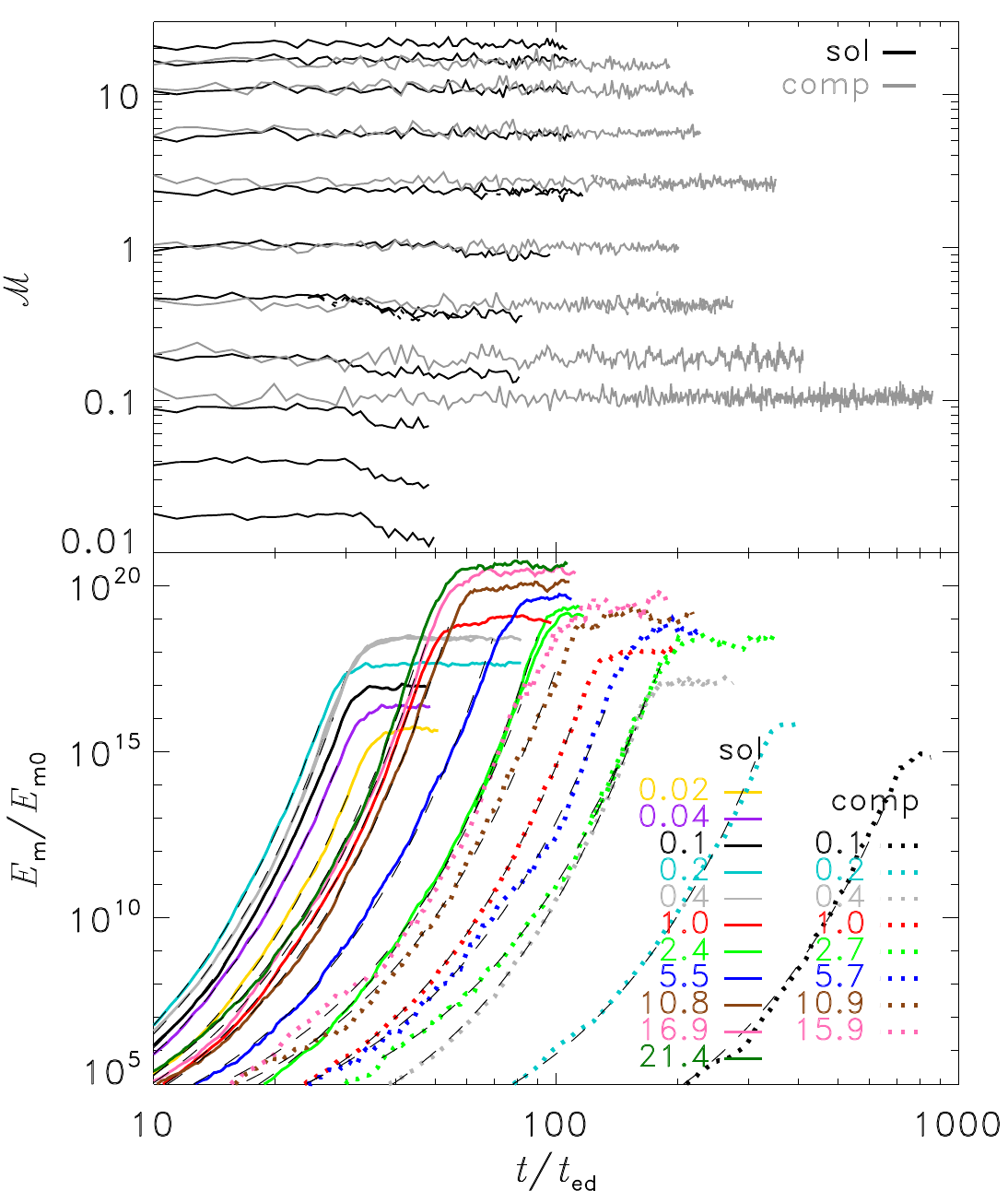}}
\caption{(Color online) Mach number, $\Ma$ (top) and magnetic energy $\Em/\Emzero$ (bottom) as a function of eddy-turnover time, $\ted$, for all runs with solenoidal (sol) and compressive (comp) forcing. The Mach number is indicated in the legend. We also add nonideal MHD models with $\Ma\approx0.4$, $2.5$ for sol.~and comp.~forcing, evolved on $256^3$, and $512^3$ grid cells. However, these models are hardly distinguishable from the corresponding ideal MHD models, because they are very similar. Thin dashed lines show fits in the exponential growth phase.}
\label{fig:evol}
\end{figure}

After an initial transient phase that lasts for $2\,\ted$, turbulence becomes fully developed and the Mach number reaches its preset value, fluctuating on a 10\% level. Figure~\ref{fig:evol} (top) shows the time evolution of $\Ma$ in all runs. Note the drop in $\Ma$ for the solenoidally driven runs with $\Ma\lesssim1$ as soon as they reach saturation. For these runs, the magnetic field has increased to a dynamically significant level, causing $\Ma$ to drop at late times, due to the back-reaction of $\bfB$ onto the flow. In contrast, in all supersonic runs and in all runs with compressive forcing, the magnetic field has little dynamical impact on the turbulent flow. Although the Mach numbers are not strongly affected in those cases, the fragmentation behavior of the gas might still change \cite{HennebelleTeyssier2008}, emphasizing the importance of magnetic fields. Figure~\ref{fig:evol} (bottom) shows that the magnetic energy grows exponentially over at least 10 orders of magnitude in each model and reaches saturation at different levels (discussed in detail below). Note that the nonideal MHD models at different resolution are almost indistinguishable from the ideal MHD models.

\begin{figure*}[t]
\includegraphics[width=0.244\linewidth]{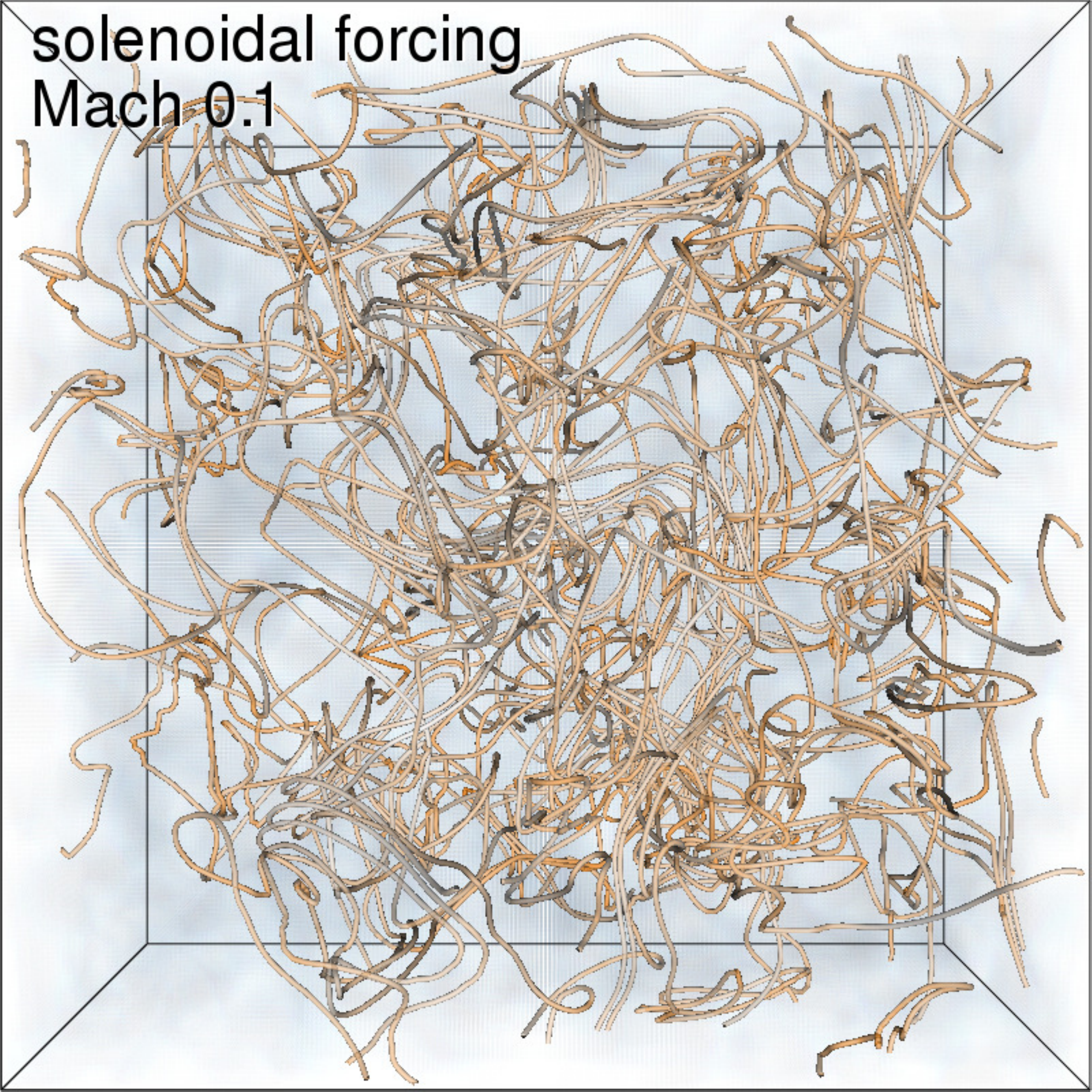}
\includegraphics[width=0.244\linewidth]{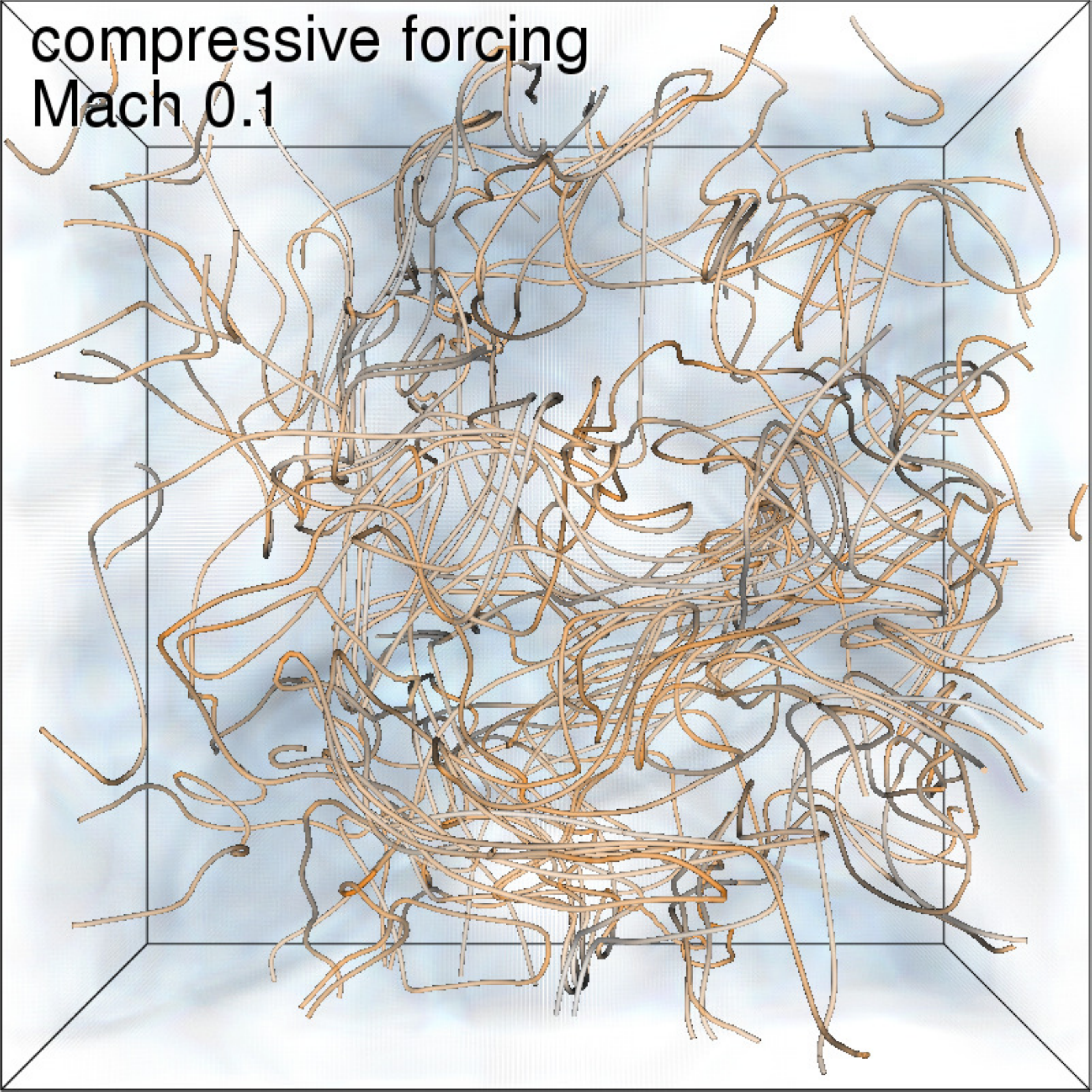}
\includegraphics[width=0.244\linewidth]{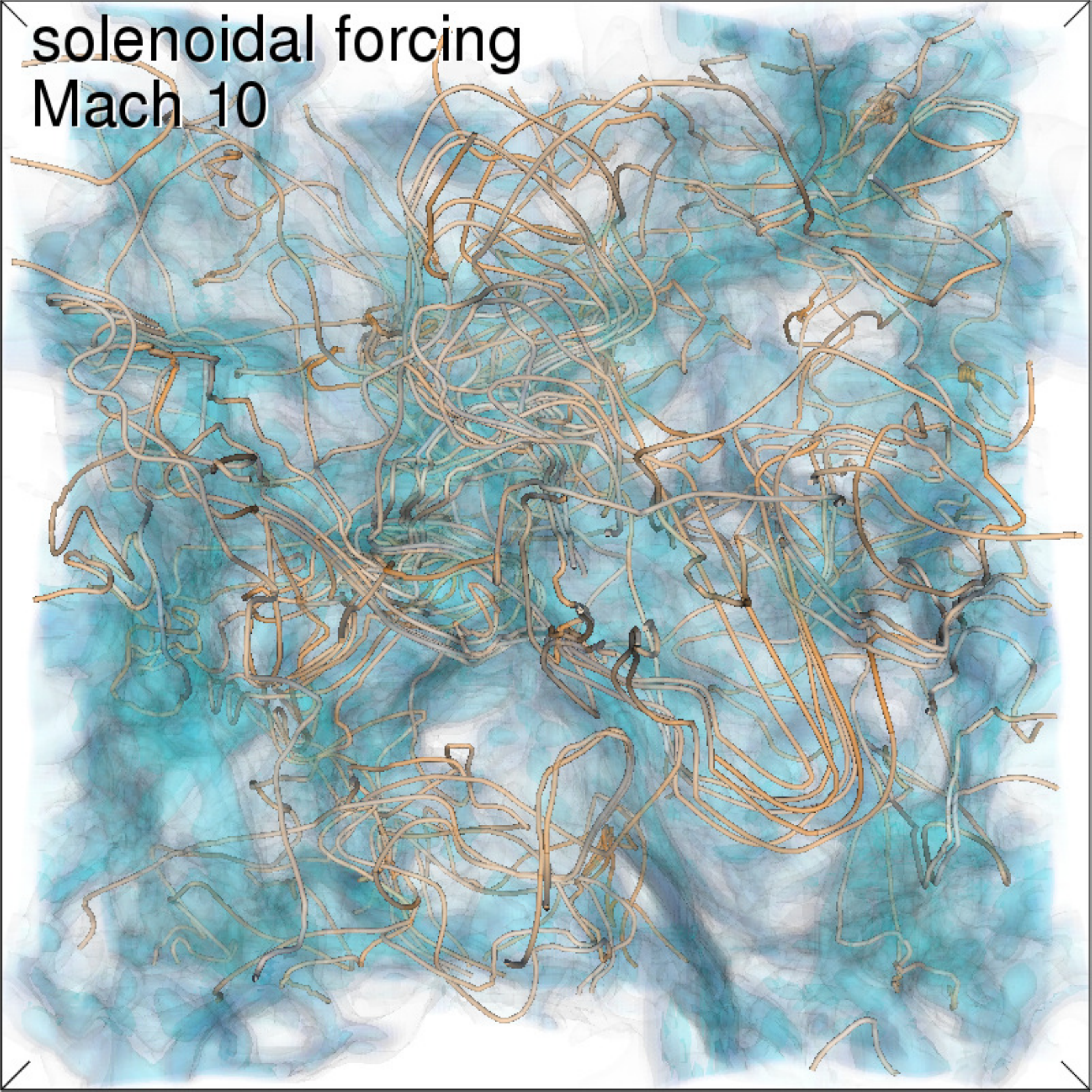}
\includegraphics[width=0.244\linewidth]{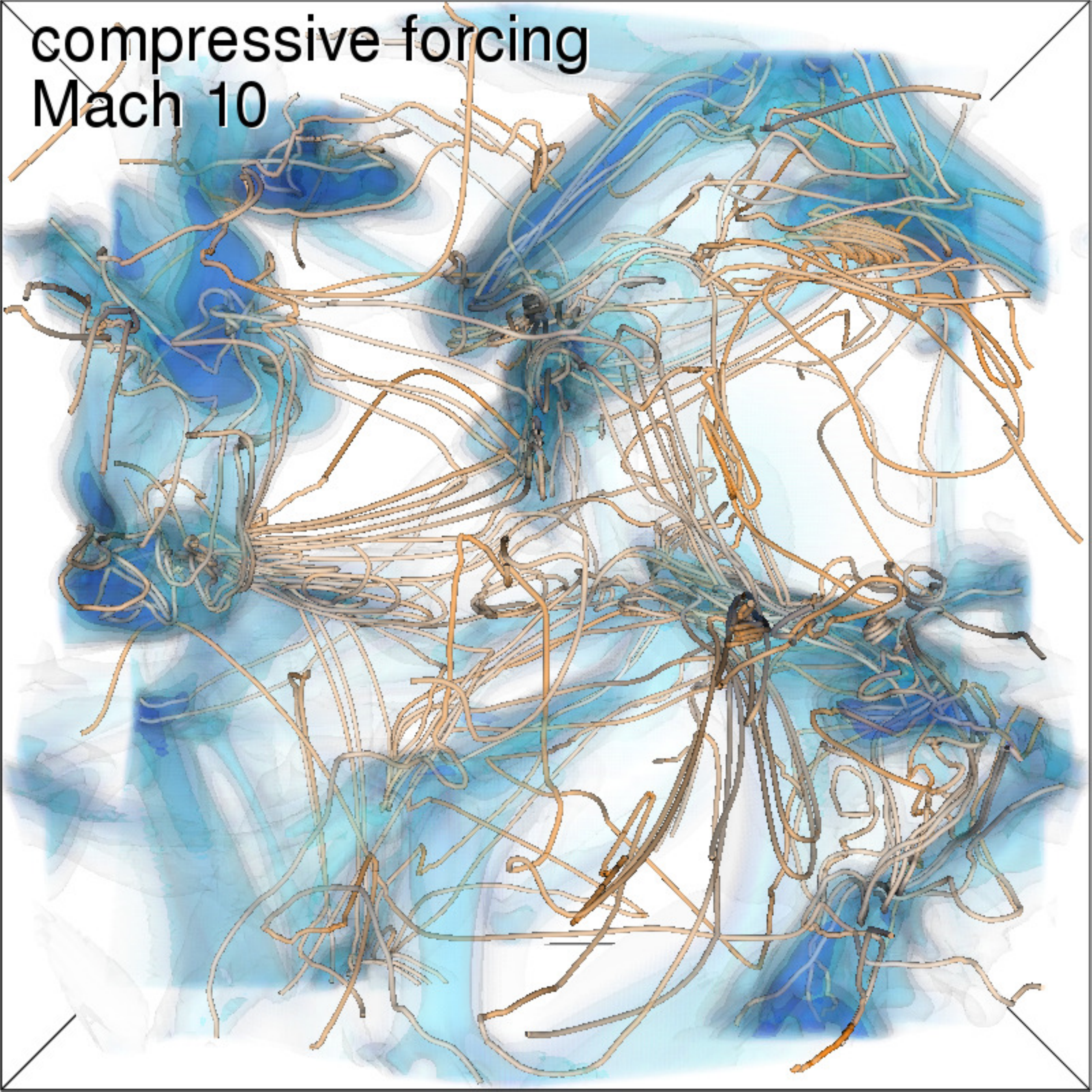}
\caption{(Color online) Three-dimensional renderings of the gas density on a logarithmic scale for $0.5\leq\rho/\rho_0\leq50$ (from white to dark blue), and magnetic field lines (orange) for solenoidal forcing at $\Ma=0.1$ (a) and $\Ma=10$ (c), and compressive forcing at $\Ma=0.1$ (b) and $\Ma=10$ (d). The stretch-twist-fold mechanism of the dynamo \cite{BrandenburgSubramanian2005} is evident in all models, but operates with different efficiency due to the varying compressibility, flow structure, and formation of shocks in the supersonic plasmas.}
\label{fig:snapshots}
\end{figure*}

\begin{figure}[t]
\centerline{\includegraphics[width=0.99\linewidth]{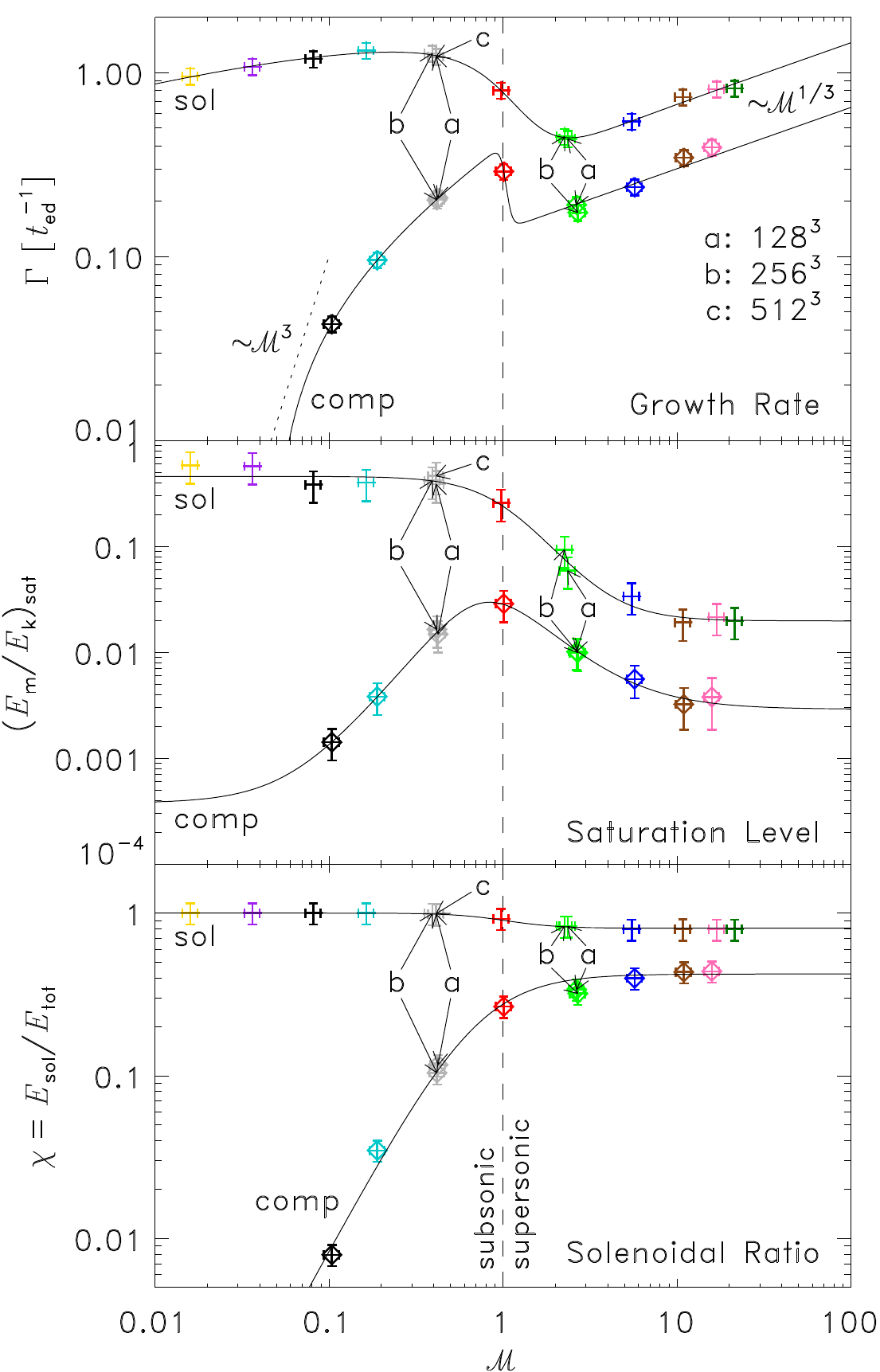}}
\caption{(Color online) Growth rate (top), saturation level (middle), and solenoidal ratio (bottom) as a function of Mach number, for all runs with solenoidal (crosses) and compressive forcing (diamonds). The solid lines show empirical fits with Eq.~(\ref{eq:fit}); see Table~\ref{tab:fit}. The arrows indicate four models ($\Ma\approx0.4$, $2.5$ for sol.~and comp.~forcing), using ideal MHD on $128^3$ grid cells (a), nonideal MHD on $256^3$ (b), and $512^3$ grid cells (c), demonstrating convergence for the given magnetic Prandtl, $\Pm\!\approx\!2$, and kinematic Reynolds number, $\mathrm{Re}\!\approx\!1500$.}
\label{fig:gratesat}
\end{figure}

\begin{table}
\caption{Parameters in Eq~(\ref{eq:fit}) for the fits in Fig.~\ref{fig:gratesat}.}
\label{tab:fit}
\begin{ruledtabular}
\begin{tabular}{lcccccc}
  & \multicolumn{2}{c}{$\Gamma\,\left[\ted^{-1}\right]$} & \multicolumn{2}{c}{$\satlev$} & \multicolumn{2}{c}{$\solratio$} \\
  & (sol) & (comp) & (sol) & (comp) & (sol) & (comp) \\
\hline
$p_0$ & -18.71                     & \phantom{-}2.251 & \phantom{-}0.020 & \phantom{-}0.037 & \phantom{-}0.808 & \phantom{-}0.423 \\
$p_1$ & \phantom{-}0.051 & \phantom{-}0.119 & \phantom{-}2.340 & \phantom{-}1.982 & \phantom{-}2.850 & \phantom{-}1.970 \\
$p_2$ & -1.059                     & -0.802                     & \phantom{-}23.33 & -0.027                    & \phantom{-}1.238 & 0 \\
$p_3$ & \phantom{-}2.921 & \phantom{-}25.53 & \phantom{-}2.340 & \phantom{-}3.601 & \phantom{-}2.850 & \phantom{-}1.970 \\
$p_4$ & \phantom{-}1.350 & \phantom{-}1.686 & 1                              & \phantom{-}0.395 & 1                              & \phantom{-}0.535 \\
$p_5$ & \phantom{-}0.313 & \phantom{-}0.139 & 0                              & \phantom{-}0.003 & 0                              & 0 \\
$p_6$ & 1/3                           & 1/3                          & 0                              & 0                              & 0                              & 0 
\end{tabular}
\end{ruledtabular}
\end{table}

Figure~\ref{fig:snapshots} shows that the high Mach number runs are dominated by shocks. Compressive forcing yields stronger density enhancements for similar Mach numbers \cite{FederrathKlessenSchmidt2008}. The magnetic field occupies large volume fractions with rather unfolded, straight field lines in the compressively driven cases, while solenoidal forcing produces more space-filling, tangled field configurations, suggesting that the dynamo is more efficiently excited with solenoidal forcing. This is quantitatively shown in Fig.~\ref{fig:gratesat} (top and middle panels), where we plot the growth rates, $\Gamma$, in the relation $\Em=\Emzero\exp(\Gamma t)$, and the saturation level, $\satlev$ with the magnetic and kinetic energies $\Em$ and $\Ek$ as a function of Mach number for all models. Both $\Gamma$ and $\satlev$ depend strongly on $\Ma$ and on the turbulent forcing. Solenoidal forcing gives growth rates and saturation levels that are always higher than in compressive forcing, as indicated by the different field geometries shown in Fig.~\ref{fig:snapshots}. Both $\Gamma$ and $\satlev$ change significantly at the transition from subsonic to supersonic turbulence. We conclude that the formation of shocks at $\Ma\approx1$ is responsible for destroying some of the coherent vortical motions necessary to drive the dynamo \cite[][]{HaugenBrandenburgMee2004}. However, as $\Ma$ is increased further, vorticity generation in oblique, colliding shocks \citep{SunTakayama2003,KritsukEtAl2007} starts to dominate over the destruction. The very small growth rates of the subsonic, compressively driven models is due to the fact that hardly any vorticity is excited. To quantify this, we plot the solenoidal ratio, i.e., the specific kinetic energy in solenoidal modes of the turbulent velocity field, divided by the total specific kinetic energy, $\chi=\solratio$ in Fig.~\ref{fig:gratesat} (bottom), which shows a strong drop of solenoidal energy for low-Mach, compressively driven turbulence. In the absence of the baroclinic term, $(1/\rho^2)\nabla\rho\times\nabla p$, the only way to generate vorticity, $\bfomega=\nabla\times\bfu$, with compressive (curl-free) forcing is via viscous interactions in the vorticity equation \cite{MeeBrandenburg2006}:
\begin{equation}
\partial_t \bfomega = \nabla\times\left(\bfu\times\bfomega\right) + \nu\nabla^2\bfomega + 2\nu\nabla\times\left(\boldsymbol{\mathcal{S}}\nabla\ln\rho\right)\,.
\label{eq:vorticity}
\end{equation}
The second term on the right hand side of the last equation is diffusive. However, even with zero initial vorticity, the last term generates vorticity via viscous interactions in the presence of logarithmic density gradients. The small seeds of vorticity generated this way are exponentially amplified by the non-linear term, $\nabla\times\left(\bfu\times\bfomega\right)$, in analogy to the induction equation for the magnetic field, if the Reynolds numbers are high enough \cite[][]{Frisch1995}. For very low Mach numbers, however, density gradients start to vanish, thus explaining the steep drop of dynamo growth in compressively driven turbulence at low Mach number. Analytic estimates \cite{MossShukurov1996} suggest that $\Gamma\propto\Ma^3$ in compressively driven, acoustic turbulence \footnote{Note that we define $\ted=L/(2\Ma\cs)$, while in \cite{MossShukurov1996}, $\ted=L/(2\cs)$, differing by a factor $\Ma$.}, indicated as dotted line in Fig.~\ref{fig:gratesat}. The solid lines are fits with an empirical model function,
\begin{equation}
\label{eq:fit}
f(x)=\left(p_0\,\frac{x^{p_1}+p_2}{x^{p_3}+p_4}+p_5\right)x^{p_6}\,.
\end{equation}
The fit parameters are given in Table~\ref{tab:fit}. We emphasize that the fits do not necessarily reflect the true asymptotic behavior of $\Gamma$ and $\satlev$. The subsonic, solenoidally driven models show very high saturation levels, $\satlev\approx40$--60\%, explaining the strong back reaction of the field, causing $\Ma$ to drop in the saturation regime (see fig.~\ref{fig:evol}, \cite{HaugenBrandenburgDobler2004}). For the growth rate, we fixed $p_6$ such that $\Gamma\propto\Ma^{1/3}$ for $\Ma\gg1$, in good agreement with our models up to $\Ma\approx20$. However, even higher $\Ma$ has to be investigated to see, if $\Gamma\propto\Ma^{1/3}$ holds in this limit. We find that $\Gamma$ depends much less on $\Ma$ in the solenoidal forcing case than in the compressive one. Nevertheless, a drop of the growth rate at $\Ma\approx1$ is noticeable in both cases. Theories based on Kolmogorov's \cite{Kolmogorov1941c} original phenomenology of incompressible, purely solenoidal turbulence predict no dependence of $\Gamma$ on $\Ma$. For instance, Subramanian \cite{Subramanian1997} derived $\Gamma=(15/24)\mathrm{Re}^{1/2}\ted^{-1}$ based on Kolmogorov-Fokker-Planck equations, in the limit of large magnetic Prandtl number, $\Pm=\nu/\eta=\Rm/\mathrm{Re}\gg1$ with the kinetic and magnetic Reynolds numbers $\mathrm{Re}=L u_\mathrm{rms}/(2\nu)$ and $\Rm=L u_\mathrm{rms}/(2\eta)$. For $\Pm\approx2$ \cite[applicable to ideal MHD, see][]{LesaffreBalbus2007}, and $\mathrm{Re}\approx1500$, corresponding to our simulations, however, we find slightly smaller growth rates, in agreement with analytic considerations \cite{BoldyrevCattaneo2004}, and with numerical simulations of incompressible turbulence for $\Pm\approx1$ \cite{SchekochihinEtAl2007,ChoEtAl2009}. Thus, an extension of dynamo theory to small $\Pm$ is needed. Moreover, extending the theory from Kolmogorov to Burgers-type, shock-dominated turbulence would be an important step forward in developing a more generalized theory of turbulent dynamos, potentially with predictive power for the supersonic regime and for compressive turbulent energy injection.

In summary, we conclude that the growth rate and saturation level of the dynamo depend sensitively on the Mach number and the energy injection mechanism of magnetized turbulence, exhibiting a characteristic drop of the growth rate at the transition from subsonic to supersonic turbulent flow. Geophysical and astrophysical dynamos operate in both, subsonic and supersonic plasmas, driven by vastly different injection mechanisms. Here we showed that strong magnetic fields are generated even in purely compressively (curl-free) driven turbulence (applicable to e.g., galactic clouds), but solenoidal (divergence-free) turbulence drives more efficient dynamos, due to the higher level of vorticity generation and the stronger tangling of the magnetic field.

\begin{acknowledgments}
{\footnotesize Stimulating discussions with A.~Brandenburg, E.~Dormy, P.~Girichidis, P.~Hennebelle, P.~Lesaffre, W.~Schmidt, and S.~Sur, and useful comments by the anonymous referees are gratefully acknowledged. C.F., G.C., and D.R.G.S.~thank for funding under the European Community's FP7/2007--2013 Grant Agreement No.~247060 and 229517. RB acknowledges funding from the DFG grant BA 3706. C.F., R.B., and R.S.K.~acknowledge subsidies from the Baden-W\"urttemberg-Stiftung (grant P-LS-SPII/18) and from the German BMBF (grant 05A09VHA). The simulations were run at the LRZ (grant pr32lo) and the JSC (grants hhd14, hhd17, hhd20). The FLASH code was in part developed by the DOE NNSA-ASC OASCR Flash Center at the University of Chicago.}
\end{acknowledgments}


%

\end{document}